\documentstyle[12pt,aaspp4,psfig,natbib]{article}
\psfigurepath{figs}   

\def\las{\mathrel{\hbox{\rlap{\hbox{\lower3pt\hbox{$\sim$}}}\hbox{\raise2pt\hbox{$<$}}}}}
\def\gas{\mathrel{\hbox{\rlap{\hbox{\lower3pt\hbox{$\sim$}}}\hbox{\raise2pt\hbox{$>$}}}}}
\newcommand{\degree}{^{\circ}}

\newcommand{\eg}{{\it e.g.\ }}
\newcommand{\ie}{{\it i.e.\ }}
\newcommand{\etc}{{\it etc.\ }}

\newcommand{\ah}{{"$\!$/hr\ }}

\begin{document}

\centerline{\bf Earth Trojan Asteroids:}
\centerline{\bf A study in support of observational searches}

\bigskip\bigskip

\author{ Paul Wiegert and Kimmo Innanen }

\affil{Dept. of Physics and Astronomy, York University, Toronto,
Ontario M3J 1P3 Canada and CITA, University of Toronto, Toronto,
Ontario M5S 3H8 Canada}

\and

\author{Seppo Mikkola}
\affil{Tuorla Observatory, University of Turku, 21500 Piikki\"o, Finland}

\begin{abstract}
Observational searches for asteroids orbiting near Earth's triangular
Lagrange points face unique obstacles. A population of such asteroids
would occupy a large projected area on the sky (possibly hundreds of
square degrees) and is not favorably placed with respect to the Sun.
Here we examine the properties of synthetic populations of Earth
``Trojans'' in order to aid in the optimization of observational
searches for them. We find that the highest on-sky projected number
densities are not located at the positions of the $L_4$ and $L_5$ points
themselves, but rather a few degrees closer to the Sun. Also,
asteroids on orbits about the $L_4$ and $L_5$ points typically brighten
as the difference between their ecliptic longitude and that of the Sun
increases owing to phase effects, but their number density on the
sky concurrently falls rapidly.
\end{abstract}

\section{Introduction} \label{section1}

General analytical solutions to the gravitational three-body problem
do not exist; however, certain special solutions do. Perhaps the best
known is the classical work of J.L. Lagrange. Given a primary $M$ with a
secondary $m \ll M$ in a circular orbit around it, he deduced that
there were five points in the system at which a massless particle
could, in a frame corotating with the secondary, remain stationary
indefinitely.  These solutions are commonly known as the Lagrange
points.  Lagrange's computations are not strictly valid for the case
of a planet orbiting in our own, more complex Solar System;
nevertheless these solutions retain much importance. Of particular
interest here are the ``tadpole'' orbits that particles may move along
in the vicinity of the two so-called ``triangular'' Lagrange
points for the case where the relative orbit of $M$ and $m$ is nearly
circular. These two $L$-points are located at the vertices of the two
equilateral triangles that can be drawn in the secondary's orbital
plane with the primary-secondary line as one side. The leading point
is usually called $L_4$, the trailing, $L_5$.  A more detailed treatment of
this subject can be found in \cite{sze67}.

Lagrange's solutions went from the realm of theoretical curiosities to
reality with the discovery in 1906 of asteroid 588 Achilles,
associated with the $L_4$ point of the Jupiter-Sun system,
\cite[]{pil79}. Since then, many more ``Trojan'' asteroids have been
found. At this writing, the Minor Planet Center lists 466 asteroids at
the $L_4$ and $L_5$ points of Jupiter and two for Mars (5261 Eureka and
1998 VF31), but as yet none has been detected near the corresponding
points of any other planet. The recent discovery \cite[]{wieinnmik97}
that asteroid 3753 Cruithne is in a 1:1 coorbital resonance (though
not a ``Trojan'' one) with Earth encourages a revisit of the
question of whether Earth has other co-orbital companions. In this
paper, we examine the properties of synthetic populations of asteroids
orbiting about the $L_4$ and $L_5$ points of Earth, for the purpose
of aiding in future observational searches. We do not investigate the
origin or stability of such objects here, but will rather assume some
population of objects on such orbits, and ask how best to detect them.

We do, however, note that the Earth/Sun mass ratio is sufficiently
small for the triangular Lagrange points to be at least linearly
stable in the 3-body context \cite[]{dan64}. Extensive investigations
of the stability of the triangular Lagrange points have been performed
by many authors, though relatively few have considered whether or not
Earth in particular can stably maintain a population of such
bodies. However, at least certain Earth Trojan asteroid (ETA) orbits
have been found (numerically) to be stable against perturbations from
the other planets in the Solar System for times up to $10^5$~yr
\cite[]{weiwet74,dun80,mikinn90}. Mikkola and Innanen (1995)
\nocite{mikinn95} studied a population of ETAs over somewhat longer
time scales ($10^6-10^7$ yr) and found instabilities at inclinations
in the range of 10--$25\degree$. Such time spans fall far short of the
$4.5\times10^9$~yr age of the Solar System. Nevertheless, even if ETA
orbits prove unstable on longer time scales, captured near-Earth
asteroids may provide a transient population of such objects.

Current telescopes and detectors should easily be able to detect ETAs,
their proximity to both Earth and the Sun making them relatively
bright. However, searches for such objects are not without their
difficulties. The nearness of these bodies to Earth means that any
Trojan group would cover a large projected area on the sky; we will
see in \S~\ref{pa:overall} that such a cloud may extend over hundreds
of square degrees with only a very broad central peak.  As well, the
necessary proximity of these asteroids to the triangular Lagrange
points requires observations to be made in morning and evening, and
hence subject to higher airmasses and the increased sky brightness of
twilight.  Observational searches for ETAs thus present interesting
challenges.

One strategy useful for the detection of faint moving bodies involves
tracking the telescope at the expected apparent angular velocity
$\dot{\theta}$ of the object in question. This results in the
searched-for object's image remaining stationary upon the detector and
building up its signal, while objects moving at different speeds (\eg
stars) can be excluded by their streak-like appearance. For this
reason, the apparent angular velocity of ETAs is examined here. One
should note that the angular velocity of an object located at either
triangular Lagrange point is not zero. Though the $L$-points are
stationary with respect to the Earth-Sun line (\ie in the frame which
corotates with the Earth), they complete a full circuit against the
background stars every year and hence their mean relative
angular rate should be $\dot{\theta} \approx 150$\ah.

The most recent search for Earth Trojan asteroids was performed by
\cite{whitho98}, who were able to cover 0.35 square degrees close to
the classical $L_4$ and $L_5$ points down to $R \sim 22.8$ using CCD
detectors. In that paper, they also provide many additional, valuable
insights affecting current search strategies. Their paper served as a
primary motivation for our work. An earlier photographic search was
performed by \cite{dunhel83}, also without detections but putting a
limit of $7 \pm 3.5$ objects with visual absolute magnitudes brighter
than $20$ ``in the $L_4$ region''. Unfortunately, they do not report
how many square degrees of the sky they sampled with the 18 plates
they obtained. It is likely to be much less than the extent of the
hypothetical Trojan clouds presented here, but in any case, the
missing information makes a comparison with other results impossible.

Though the absence of detections may indicate an absence of ETAs,
searches to date have likely only sampled a relatively small portion
of the potential Trojan cloud, and we will see in \S~\ref{pa:results}
that the existence of a population of Earth Trojans cannot be rejected
at this time.

Our model is presented in \S~\ref{pa:model}; the results are discussed
in \S~\ref{pa:results} and our conclusions are presented in
\S~\ref{pa:conc}.

\section{Model calculations} \label{pa:model}

In order to determine how best to detect a population of Earth
Trojans, something must first be known about their properties. Studies
of main-belt and near-Earth asteroids allow us to make estimates of
ETA albedos and other surface properties.  Estimating their spatial
distribution is less straightforward, but our knowledge of the Jovian
Trojan group and current numerical models of the Solar System
should provide us with a reasonable starting place.

We examine two synthetic populations of Earth Trojans. Both have
semi-major axes chosen uniformly and randomly in the coorbital region
between the $L_2$ and $L_3$ points, \ie $0.997 \le a \le 1.003$~AU,
based on the stability limits for Earth Trojans found by
\cite{weiwet74}. The width of this zone matches the width of the
resonance overlap zone predicted by the $\mu^{2/7}$ law
\cite[]{wis80,dunquitre89,murhol98}. This relation states that, if $a$
is the semi-major axis of a planet and $\mu$ is the ratio of its mass
to the total (ie. planet plus Sun), then small bodies with semimajor
axes differing from $a$ by $\Delta a \gas \mu^{2/7}$ are isolated from
the 1:1 resonance by a region of quasi-periodic (\ie non-chaotic)
trajectories. The $\mu^{2/7}$ law thus defines a boundary beyond which
a simulated test body has effectively escaped from 1:1 resonance.

Apart from the value of the semi-major axis, the first (or ``broad'')
population is given orbital elements which are chosen based on the
breadth of the distribution of Jupiter's Trojans, as listed on the
Minor Planet Center's web site
(http://cfa-www.harvard.edu/iau/lists/Trojans.html).  The
eccentricities $e$ and inclinations $i$ are chosen randomly from the
uniform distributions $0 \le e \le 0.3$ and $0\degree \le i \le
30\degree$. A random longitude of the ascending node $\Omega$ and
argument of perihelion $\omega$ is assigned to each particle, and its
mean anomaly chosen such that its mean longitude is either $60\degree$
ahead of or behind the Earth.

A second (or ``narrow'') population is also examined, its element
distributions based on the locations of the peaks in the orbital
element distributions of Jupiter's Trojans.  In this case, the
eccentricity and inclination are chosen from $0 \le e \le 0.1$ and
$0\degree \le i \le 10\degree$ instead.

The particle populations chosen are then integrated for $10^5$ years
with a Wisdom-Holman \cite[]{wishol91} integrator using a 3-day time
step in a Solar System including all the planets except Pluto. Those
particles which remain on Trojan (or tadpole) orbits throughout the
entire integration, (\ie those whose longitude relative to Earth
remains either consistently ahead of or behind our planet) will be
considered our sample. This integration interval allows the removal of
objects which do not remain on Trojan orbits (with the tacit
assumption that such short-lived objects do not constitute an
appreciable fraction of the Earth Trojan population). Of the broad
population, 63 of 100 initial conditions pass this criterion
(primarily those at low $e$); 95 of 100 of the narrow population's
members do. We have also performed $10^6$~yr integrations for the
trailing Lagrange point with little change in the survival fraction or
other results.

We have deliberately chosen to exclude horseshoe orbits from this
study, though asteroid 3753 Cruithne \cite[]{wieinnmik97} is proof
that bodies on non-tadpole 1:1 mean motion resonances exist.
Nevertheless, the apparent lack of such a population for Jupiter has
lead us to examine the possible traditional Trojan population first,
though a similar study for possible horseshoe orbits would also be of
interest.

During the simulations, the eccentricities and inclinations of the
particles are monitored. Various perturbations induce oscillations in
these quantities, but their amplitudes are rather small. For the
eccentricity, $e$ is found to vary by $\pm 0.01$ near $e=0$ up to $\pm
0.025$ at $e \gas 0.2$. The inclination varies typically by $\pm
3\degree$.  All of $e$, $i$ and $\omega$ show these oscillations with
periods of $10^4-10^5$~yr, typical of secular perturbations by the
giant planets. The longitude of the node varies on
much shorter time scales ($10^2$~yr), most likely driven by close
encounters with the planet, which have a similar time scale.

In addition to the orbital elements, the apparent geocentric position
and other properties (phase, apparent magnitude, \etc) are measured at
regular (but non-commensurable) intervals. This set of synthetic
``observations'' should provide a benchmark with which to optimize
observing strategies. 

The apparent visual magnitude of our simulated asteroids is calculated
using the IAU two-parameter magnitude relation \cite[]{bowhapdom89},
which we will term $M_V(r_{\odot}, r_{\oplus}, \alpha, H, G)$;
$r_{\odot}$ is the asteroid-Sun distance, $r_{\oplus}$ is the
asteroid-Earth distance and $\alpha$ is the phase angle \ie the angle
between the Sun and Earth as seen from the asteroid.  The two
parameters in question are the absolute magnitude $H$ of the asteroid,
and $G$, the so-called `slope parameter', which describes the change
in brightness with phase.  These parameters depend on the particular
properties of a given asteroid, and thus are unknown for our
sample. However, since apparent magnitude scales linearly with $H$, we
simply calculate the difference between the apparent and absolute
magnitude, which we will call the magnitude adjustment $\Delta M_V$,
defined such that
\begin{equation}
\Delta M_V(r_{\odot},r_{\oplus},\alpha,G) = 
M_{V}(r_{\odot},r_{\oplus},\alpha,H,G) - H.
\label{eq:magadj}
\end{equation}
The $G$ factor has a relatively weak influence on apparent magnitude,
and we simply adopt the value of 0.15, noting that the most common
asteroids have typical $G$ values ranging from 0.1 (C-type) to 0.25
(S-type) \cite[]{bowhapdom89,piimaglag97}.  Note that no albedo or
diameter need be assumed, these unknowns being subsumed into the
absolute magnitude $H$.

The detection of faint ETAs will be hampered by the presence of
crowded star fields; in particular, the positions of the Lagrange
points relative to the Milky Way should be considered. This issue was
addressed by \cite{whitho98} and we simply quote it here for
completeness. The $L_4$ point is at its highest galactic latitudes
during April through June and October through December; the $L_5$ point,
during December through February and June through August.

Asteroids orbiting both the leading and trailing triangular Lagrange
points have been studied.  The populations of these two regions,
should they exist, may differ (\eg the leading/trailing ratio for
Jupiter Trojans is $292/176 \approx 1.66$, though observational biases
complicate this picture), but our dynamical simulations, as yet, give
no hint of such a possible skew. In both leading and trailing cases,
almost the same fraction of the ETAs (63\% and 95\% for the broad and
narrow populations respectively) remain on tadpole orbits for the full
$10^5$~yr integration, and the resulting distributions of elements,
positions, \etc are very similar, though the randomized initial
conditions ensure that the leading and trailing distributions are not
identical. In the interests of brevity, we will discuss here only the
trailing distribution, with the afore-mentioned caveat in mind.

Three questions will spearhead our examination of our simulated Earth
Trojan population. First, where is the greatest projected density of
ETAs on the sky?  Second, where are ETAs at their brightest as
seen from Earth? And third, which ETAs can be observed under the
darkest skies?

\section{Results} \label{pa:results}

\subsection{Overall sample} \label{pa:overall}

Let the geocentric ecliptic latitude of an ETA at any given time be
$\beta$; and let the ecliptic longitude relative to the Sun be
$\ell$. Then, the Sun is always at $\beta = 0\degree, \ell = 0
\degree$. Figure~1 displays the distribution of simulated
observations (as described above) against the axes of $\beta$ and
$\ell$ for both the broad and narrow populations. On this and on the
following plots, the darker pixels indicate those at which more
synthetic observations where made, while the lighter ones indicate a
relative absence\footnote{Note: pixel counts are not weighted by
apparent brightness, thus these plots are not strictly comparable to
long time exposure ``images''.}. The contours are labeled according to
the total number of observations interior to them, and account for the
(usually small) number of data points outside the presented
frame. Thus the contour labeled ``10'' is that interior to which the
most densely packed 10\% of the observations fall; similarly, the
``25'' contour includes 25\% of all observations, and so on. Various
statistics of the ETA distributions are presented in Table~I.


The plots were subjected to a 3x3 boxcar filter to smooth the
contours. The (pre-smoothing) noise-related uncertainty in each pixel
is small in the most heavily populated regions of the graph, with
typical peak values of 200 counts, a noise level of roughly
7\%. However, the observations are not all independent (and thus not
strictly Poissonian in their noise characteristics), belonging to the
trajectories of only $\sim 100$ test particles and thus effectively
increasing the uncertainty in the results. This, together with the
assumptions made about the ETAs orbital elements, implies that
small-scale structures in the figures must be interpreted with
particular caution. Indeed here we will examine only the broad
features of these distributions. The one exception we make is to
present a magnified view of the peak of the on-sky
distribution. Frames $(a)$ and $(b)$ of Figure~1 might conceal a
narrow central peak or similar feature of observational
interest. However, the high magnification frames $(c,d)$ make it clear
that there is no particularly overdense region in the vicinity of the
Lagrange points.  The reader should note that the details of the
contours at low and high magnifications differ somewhat owing to the
different binning, which also affects the smoothing algorithm.  Both
Figs.~$1c$ and $1d$ have peak pixel values $\sim 100$ (RMS
noise $\sim10$\%).

Figure~1 makes clear one of the key difficulties facing
ETA searchers: the wide area of sky to be searched. As an
illustration, the 10\% contour, though it encompasses roughly 60 and
25 square degrees for the broad and narrow populations respectively,
is only occupied by 10\% of the ETA population at any given
time. \cite{whitho98} placed an upper limit of 3 objects larger than a
few hundred meters per square degree in the vicinity of the $L_4$ and
$L_5$ points. Assuming the 10\% contour is populated at this level,
75-180 objects this size could be in this region at any instant. Thus,
though past searches in this region have failed to discover such
asteroids, the existence of ETAs is still far from being ruled out.

The distribution of simulated ETAs is centered on zero ecliptic
latitude, expected from the symmetric distribution of the inclinations
and the uniform distribution of the other angular elements. It is
highest within a degree or two of the ecliptic plane and has a much
broader distribution along the ecliptic. 

One should note, however, that the centroid of the highest density
contour is not at $60\degree$ longitude \ie is not in the direction of
the Lagrange point. Rather the centroid is slightly interior to this
point at $\ell \sim 55\degree$.  Since a constant distribution of
asteroids along Earth's orbit would remain constant when projected
on to the sky, we conclude that the displacement of the centroid is
primarily due to the asteroids spending more time in the elongated
``tail'' of the tadpole, rather than to purely geometrical projection
effects. The density variation between $55\degree$ and $60 \degree$ is
weak ($\sim 25$\%) and would make little difference to a survey which
could sample only a small area (\eg less than 1 sq. degree).
Nevertheless, this factor should be taken into account when wide-field
surveys are attempted.

Figure~2 presents the distribution of the simulated
asteroids' apparent magnitude adjustment $\Delta M_V$ versus ecliptic
latitude and longitude. Asteroids near the projected positions of the
Lagrange points are typically 2 to 2.5 magnitudes fainter than their
absolute magnitude \eg a 6-km diameter C-type asteroid ($H \approx
15$, Bowell and Lumme 1979; Rabinowitz 1993) \nocite{bowlum79,rab93}
would have an apparent magnitude $M_V \approx 17-17.5$. If ETAs have
a size distribution similar to NEAs \cite[eg. $N(H) \propto e^{0.9H}$;
]{rab94} then, given that current limits allow a population of perhaps
$10^3$ objects with diameters greater than 100 meters ($H \approx
23$), one might expect $\sim 10$ objects in the 5 kilometer range. Of
course, such numbers are based on rather loose upper limits, but the
existence of kilometer-sized ETAs cannot yet be ruled out. Such
objects would certainly be detectable by modern facilities, though
again questions of sky brightnesses and tracking rates complicate the
issue significantly.


The brightest simulated observations are outside the plot area of
Figs.~$2a$ and $b$. The distribution continues up to
brighter magnitudes at ecliptic longitudes near $90\degree$, though at
very low number densities. These brightest data points are caused by
objects passing very close to Earth. Asteroids near to, but slightly
outside Earth's orbit, have their brightness enhanced both by their
proximity and their low phase angle. These brightest observations will
be discussed in more detail in \S~\ref{pa:bright}.

Figure~3 examines the distributions of magnitude versus
angular velocity $\dot{\theta}$. The ETAs cluster around 150\ah, but
with significant scatter: only 25\% and 50\% of the broad and narrow
populations respectively have 140\ah$ < \dot{\theta} <
160$\ah. Figure~3 also displays the components of
$\dot{\theta}$ both parallel ($XY$) and perpendicular ($Z$) to the
ecliptic. The $XY$ component is centered on $\dot{\theta} \approx
150$\ah, that being the angular velocity of the Lagrange points. The
$Z$ component, which can be significant, is a result of the
inclinations of the assumed ETAs. We concur with \cite{whitho98} that
tracking the telescope at at rate of 150\ah along the ecliptic is
appropriate but will reduce one's sensitivity to high-$i$ objects.

We should point out that non-Trojan near-Earth asteroids could pass
through the regions of the sky considered here with apparent motions
similar to those of Trojans. Orbital determinations would thus be
necessary to confirm the nature of any such detections made.


\subsection{Brightest observations} \label{pa:bright}

The brightest ETAs have an obvious advantage for detection. 
Here we will define the sample of the brightest ETA observations
as those with $\Delta M_V < 1.5$. Roughly 5\% of the simulated
observations fall into this category. Figure~4 presents
their distribution on the sky and versus magnitude and apparent
angular velocity. Some general statistics of the brightest
observations are listed in Table~II.


The brightest simulated observations are found to be clustered
somewhat differently than the general sample.  The brightest
observations tend to occur nearer to Earth, both because of the
$1/r^2$ effect and because of decreased phase angles for bodies
outside Earth's orbit.  The proximity of these observations to us
results occasionally in high ecliptic latitudes (up to $\beta \gas \pm
80\degree$ for the broad population). The increased brightness
resulting from decreased phase angle favors asteroids with positions
outside Earth's orbit, and results in a shift of the center of the
on-sky distribution to larger ecliptic longitudes. The sample of the
brightest observations, as defined above, clusters at $\ell \gas
72\degree$, but this particular longitude value is an artifact of our
choice of cutoff in $\Delta M_V$ and the brightening trend for the
observations at high $\ell$ (\ie low phase). Had we chosen $\Delta M_V
\le 1$ or $\le 2$, we would have found the distribution cluster around
$78\degree$ or $65\degree$ respectively instead.

We conclude that searches for ETAs at high ecliptic longitudes are
aided by the brightening of the asteroids (and the darker skies
available), though at a significant cost in on-sky number density.

\subsection{Large ecliptic longitudes} \label{pa:farthest}

Astronomical observations of objects as faint as ETAs require
reasonably dark skies. However, by virtue of their position relative
to the Earth-Sun line, the Lagrange points cross the meridian in the
daytime. Of course, observations can be obtained of these regions in
the evenings and mornings, but are hampered by twilight and high
airmasses.  

The difference between the right ascension of the Sun and the object
in question provides a measure on the time between the setting and
rising of the two. However, RA is a function of the angle between the
Earth-Sun line and Earth's axis, and thus a function of the time
of year. The question of observability is also affected by the
position on the globe of the observatory.

Instead of performing a detailed examination of observability for
particular observatories over the course of a year, we use a cruder
and more general criterion for determining simulated observations
observability. We take the ecliptic longitude difference $\ell$ of
an ETA as a measure of the object's apparent distance from the Sun
(\ie we take Earth's obliquity to be zero). The solar elongation
(\ie the angle between the Sun and the object as seen from Earth) is
not a suitable measure here, for we are interested primarily in the
Sun-asteroid angular separation {\it perpendicular} to Earth's
terminator.

We examine the set of ETA observations with longitude differences
$\ell \ge 65 \degree$.  Roughly 18\% of our data points fall into this
category, for both broad and narrow populations. The results are
displayed in Fig.~5 and Table~III. Though
scattered observations extend out beyond $120\degree$ longitude, only
$\sim 1$\% of observations for each population are at $\ell >
80\degree$. These data points are typically brighter than average
owing to the decreased phase angle at high longitude, reflected in
Table~III by a $\Delta M_V \approx 1.8$, compared to 2.2
for the sample as a whole (Table~I).


\section{Conclusions} \label{pa:conc}

We have examined two hypothetical populations of Earth Trojan
asteroids with an eye to clarifying the observational strategies
necessary to detect them efficiently. Synthetic observations of an ETA
cloud are clustered around the triangular Lagrange points, as
expected, and may cover hundreds of square degrees of sky. As a
result, current observational limits still allow populations of Earth
Trojans of several hundred objects larger than a few hundred meters in
size.

However, pointing one's telescope directly towards one of Earth's
Lagrange points does not result in an optimal search of the region of
the sky most heavily populated by ETAs. In fact, a slightly higher
concentration can be found at smaller ecliptic longitude difference
$\ell \approx 55\degree$. Also, there is a trend for ETAs to brighten
as $\ell$ increases due to the decreasing phase angle, and searches
also benefit from the darker skies these positions allow.  However,
the density of objects decreases sharply with $\ell$. Thus,
observational searches should perhaps be tailored to optimize the
local strengths, be they wide field of view or faint limiting
magnitude.

Acknowledgements: We gratefully thank D. Tholen and R. Whiteley for
helpful comments, and S. J. Bus and an anonymous referee for their
valuable insights. This work was performed while PW was a National
Fellow at the Canadian Institute for Theoretical Astrophysics, and was
supported in part by the Natural Sciences and Engineering Research
Council of Canada. KI expresses his thanks to Prof. Mauri Valtonen of
the Tuorla Observatory for his traditional generous hospitality during
the preparation of this manuscript.

Note: if this preprint has been obtained via the LANL preprint
archive, the numbering of the contours in the figures may be dificult
to read owing to the file size restrictions imposed by the archive. A
version of this preprint with full-resolution figures is available at
http://aries.phys.yorku.ca/$\sim$wiegert/preprints.html.

\bibliographystyle{icarusbib} \bibliography{Wiegert}

\begin{thebibliography}{Weissman and Wetherill(1974)}

\bibitem[Bowell and Lumme(1979)]{bowlum79}
Bowell, E. and K.~Lumme.
\newblock 1979.
\newblock Colorimetry and magnitudes of asteroids.
\newblock In {\em Asteroids}, (T.~Gehrels, Ed.), pp. 132--169. University of
  Arizona Press, Tucson.

\bibitem[Bowell {\it et~al.}(1989)]{bowhapdom89}
Bowell, E., B.~Hapke, D.~Domingue, K.~Lumme, J.~Peltoniemi,  and A.~Harris.
\newblock 1989.
\newblock Application of photometric model to asteroids.
\newblock In {\em Asteroids II}, (R.~Binzel, T.~Gehrels, and M.~Matthews,
  Eds.), pp. 524--556. University of Arizona Press, Tucson.

\bibitem[Danby(1964)]{dan64}
Danby, J. M.~A.
\newblock 1964.
\newblock Stability of the triangular {Lagrange} points in the elliptic
  restricted problem of three bodies.
\newblock {\em Astron. J.} {\bf 69}, 165--172.

\bibitem[Dunbar and Helin(1983)]{dunhel83}
Dunbar, R.~S. and E.~F. Helin.
\newblock 1983.
\newblock Estimation of an upper limit on the {Earth} {Trojan} asteroid
  population from {Schmidt} survey plates.
\newblock {\em Bull. Amer. Astron. Soc.} {\bf 15}, 830.

\bibitem[Dunbar(1980)]{dun80}
Dunbar, R.~S.
\newblock 1980.
\newblock {\em Dynamics and stability of {Trojan} librations in the
  {Earth}-{Sun} system: implications for the existence of an {Earth} {Trojan}
  asteroid group}.
\newblock PhD thesis, Princeton.

\bibitem[Duncan {\it et~al.}(1989)]{dunquitre89}
Duncan, M., T.~Quinn,  and S.~Tremaine.
\newblock 1989.
\newblock The long-term evolution of orbits in the {Solar} {System}: a mapping
  approach.
\newblock {\em Icarus} {\bf 82}, 402--418.

\bibitem[Mikkola and Innanen(1990)]{mikinn90}
Mikkola, S. and K.~Innanen.
\newblock 1990.
\newblock Studies of {Solar} {System} dynamics {II}. {The} stability of
  {Earth}'s {Trojans}.
\newblock {\em Astron. J.} {\bf 100}, 290--293.

\bibitem[Mikkola and Innanen(1995)]{mikinn95}
Mikkola, S. and K.~Innanen.
\newblock 1995.
\newblock On the stability of high inclination {Trojans}.
\newblock {\em Earth, Moon, and Planets} {\bf 71}, 195--198.

\bibitem[Murray and Holman(1998)]{murhol98}
Murray, N. and M.~Holman.
\newblock 1998.
\newblock Diffusive chaos in the outer asteroid belt.
\newblock {\em Astron. J.} {\bf 114}, 1246--1259.

\bibitem[{Piironen} {\it et~al.}(1997)]{piimaglag97}
{Piironen}, J., P.~{Magnusson}, C.~I. {Lagerkvist}, I.~P. {Williams}, M.~E.
  {Buontempo},  and L.~V. {Morrison}.
\newblock 1997.
\newblock Physical studies of asteroids. {XXXI}. {Asteroid} photometric
  observations with the {Carlsberg} {Automatic} {Meridian} {Circle}.
\newblock {\em Astron. Astrophys. Suppl.} {\bf 121}, 489--497.

\bibitem[Pilcher(1979)]{pil79}
Pilcher, F.
\newblock 1979.
\newblock Circumstances of minor planet discovery.
\newblock In {\em Asteroids}, (T.~Gehrels, Ed.), pp. 1130--1154. University of
  Arizona Press, Tucson.

\bibitem[Rabinowitz(1993)]{rab93}
Rabinowitz, D.~L.
\newblock 1993.
\newblock The size distribution of earth-approaching asteroids.
\newblock {\em Astrophys. J.} {\bf 407}, 412--427.

\bibitem[{Rabinowitz}(1994)]{rab94}
{Rabinowitz}, D.~L.
\newblock 1994.
\newblock The size and shape of the near-earth asteroid belt.
\newblock {\em Icarus} {\bf 111}, 364--377.

\bibitem[Szebehely(1967)]{sze67}
Szebehely, V.
\newblock 1967.
\newblock {\em Theory of Orbits}.
\newblock Academic Press, New York.

\bibitem[Weissman and Wetherill(1974)]{weiwet74}
Weissman, P.~R. and G.~W. Wetherill.
\newblock 1974.
\newblock Periodic {Trojan}-type orbits in the {Earth}-{Sun} system.
\newblock {\em Astron. J.} {\bf 79}, 404--412.

\bibitem[Whiteley and Tholen(1998)]{whitho98}
Whiteley, R.~J. and D.~J. Tholen.
\newblock 1998.
\newblock A {CCD} search for {Lagrangian} asteroids in the {Earth}-{Sun}
  system.
\newblock {\em Icarus} {\bf 136}, 154--167.

\bibitem[Wiegert {\it et~al.}(1997)]{wieinnmik97}
Wiegert, P., K.~Innanen,  and S.~Mikkola.
\newblock 1997.
\newblock An asteroidal companion to the {Earth}.
\newblock {\em Nature} {\bf 387}, 685--686.

\bibitem[Wisdom and Holman(1991)]{wishol91}
Wisdom, J. and M.~Holman.
\newblock 1991.
\newblock Symplectic maps for the n-body problem.
\newblock {\em Astron. J.} {\bf 102}, 2022--2029.

\bibitem[Wisdom(1980)]{wis80}
Wisdom, J.
\newblock 1980.
\newblock The resonance overlap criterion and the onset of stochastic behavior
  in the restricted three body problem.
\newblock {\em Astron. J.} {\bf 85}, 1122--1133.

\end{thebibliography}
\newpage

\vspace*{2in}
\centerline{\begin{tabular}{l|cccc|cccc}\hline
                & \multicolumn{4}{c|}{Broad} & \multicolumn{4}{c}{Narrow} \\
                     &  Min.& Max.& Med.& Mean& Min.& Max.& Med.& Mean \\ \hline \hline
Ecl. Latitude (deg)  & -85  & 79  & 0.0 & 0.0 & -26 & 26  & 0.0 & 0.0 \\
Ecl. Longitude (deg)& 2    & 126 & 54  & 53  & 4   & 90  & 56  & 53  \\
Mag. Adjustment (mag)& -0.6 & 3.1 & 2.2 & 2.2 & 0.1 & 2.8 & 2.3 & 2.2 \\
Phase (deg)          & 3    & 120 & 55  & 53  & 5   & 92  & 56  & 53  \\
Dist. from Earth (AU) &  0.2 & 2.1 & 1.1 & 1.2 & 0.3 & 2.1 & 1.1 & 1.2 \\
Angular vel.(\ah)    &  71  & 430 & 147 & 148 & 116 & 209 & 148 & 148 \\ \hline
\end{tabular}}
\bigskip\bigskip
\noindent{Table I: Statistics concerning the distribution of the broad
and narrow simulated Earth Trojan population.}

\newpage

\vspace*{2in}
\centerline{\begin{tabular}{l|cccc|cccc}\hline
                  & \multicolumn{4}{c|}{Broad} & \multicolumn{4}{c}{Narrow} \\
                      & Min.& Max. & Med. & Mean & Min.& Max.& Med. & Mean \\ \hline \hline
Ecl. Latitude (deg)   & -85 & 79   & -0.3 & -1.0 & -26 & 26  & -0.1 & -0.2 \\
Ecl. Longitude (deg) & 29  & 126  & 72   & 73   & 61  & 90  & 74   & 74   \\
Mag. Adjustment (mag) & -0.6& 1.5  & 1.3  & 1.2  & 0.1 & 1.5 & 1.3  & 1.2  \\
Phase (deg)           & 54  & 113  & 74   & 75   & 63  & 92  & 75   & 74   \\
Dist. from Earth (AU)     & 0.2 & 0.65 & 0.55 & 0.53 & 0.3 & 0.65& 0.55 & 0.54 \\
Angular vel.(\ah)     & 71  & 430  & 166  & 169  & 121 & 209 & 154  & 155  \\ \hline
\end{tabular}}
\bigskip\bigskip
\noindent Table II: Statistics concerning the distribution of the brightest
$(\Delta M_V < 1.5)$ members of the simulated Earth Trojan
population.

\newpage

\vspace*{2in}
\centerline{\begin{tabular}{l|cccc|cccc}\hline
                  & \multicolumn{4}{c|}{Broad} & \multicolumn{4}{c}{Narrow} \\
                      & Min.& Max. & Med. & Mean & Min.& Max.& Med. & Mean \\ \hline \hline
Ecl. Latitude (deg)   & -80 & 79   & -0.23& -0.14& -26 & 26  & 0.0 & 0.0 \\
Ecl. Longitude (deg) & 65  & 126  & 70   & 71   & 65  & 90  & 69   & 70   \\
Mag. Adjustment (mag) & -0.6& 2.7  & 1.8  & 1.8  & 0.1 & 2.4 & 1.8  & 1.7  \\
Phase (deg)           & 48  & 97   & 65   & 65   & 63  & 92  & 75   & 74   \\
Dist. from Earth (AU)     & 0.2 & 1.22 & 0.76 & 0.76 & 0.29& 1.1 & 0.75 & 0.73 \\
Angular vel.(\ah)     & 82  & 407  & 146  & 148  & 116 & 209 & 147  & 148  \\ \hline
\end{tabular}}
\bigskip\bigskip
\noindent Table III: Statistics concerning the distribution of the
members of the simulated Earth Trojan population with the greatest
longitudes ($\ell \ge 65\degree$).
%


\end{document}